\documentclass[twocolumn,floats,showpacs,pra]{revtex4}
\usepackage[above]{placeins}
\usepackage{amsmath}
\usepackage{graphicx}

\begin{document}

\title{Magnetic monopole and the nature of the static magnetic field}
\author{Xiuqing Huang$^{1,2}$}
\email{xqhuang@nju.edu.cn}
\affiliation{$^1$Department of Physics and National Laboratory of Solid State
Microstructure, Nanjing University, Nanjing 210093, China \\
$^{2}$ Department of Telecommunications Engineering ICE, PLA University of
Science and Technology, Nanjing 210016, China}
\date{\today}

\begin{abstract}
We investigate the factuality of the hypothetical magnetic monopole and the
nature of the static magnetic field. It is shown from many aspects that the
concept of the massive magnetic monopoles clearly is physically untrue. We
argue that the static magnetic field of a bar magnet, in fact, is the static
electric field of the periodically quasi-one-dimensional electric-dipole
superlattice, which can be well established in some transition metals with
the localized $d$-electron. This research may shed light on the perfect
unification of magnetic and electrical phenomena.
\end{abstract}

\maketitle

\section{Introduction}

The general concept of symmetry plays an important role in physics and other
fields of science. It is well known that the behavior of electric and
magnetic fields can be completely described by Maxwell's equations. For
time-varying fields, the differential form of these four important equations
in cgs (short for centimeter, gram, second) is given by
\begin{eqnarray}
\nabla \cdot \mathbf{E} &=&4\pi \rho _{e},  \notag \\
\nabla \cdot \mathbf{B} &=&0,  \notag \\
\nabla \times \mathbf{E} &=&-\frac{1}{c}\frac{\partial \mathbf{B}}{\partial t%
},  \notag \\
\nabla \times \mathbf{B} &=&\frac{1}{c}\frac{\partial \mathbf{E}}{\partial t}%
+\frac{4\pi }{c}\mathbf{J}_{e},  \label{maxwell}
\end{eqnarray}%
where \textbf{E} is the electric field, \textbf{B} is the magnetic field, $%
\rho _{e}$ is the electric charge density, $\mathbf{J}_{e}$ is the electric
current density and $c$ is the speed of light in a vacuum.

Without electromagnetic sources ($\rho _{e}=0$; $\mathbf{J}_{e}=0$), we can
see clearly that the set of Eq. (\ref{maxwell}) will remain invariant under
the following duality transformations
\begin{equation}
\mathbf{E}\rightarrow \mathbf{B};\quad \mathbf{B}\rightarrow -\mathbf{E}.
\label{duality}
\end{equation}

\begin{figure}[tbp]
\begin{center}
\resizebox{1\columnwidth}{!}{
\includegraphics{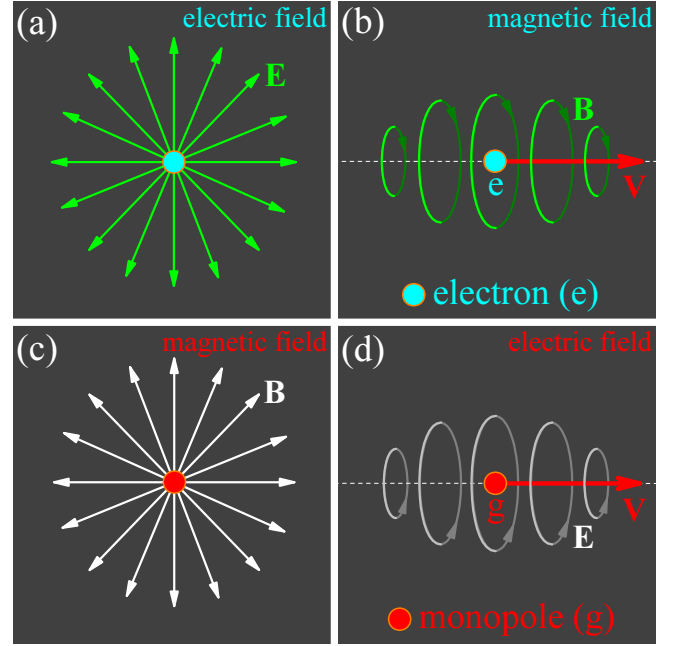}}
\end{center}
\caption{Electric (\textbf{E}) and magnetic (\textbf{B}) field lines
generated by electron (or monopole) and by their motion with velocity
\textbf{V}. (a) The electric field of a static electron with electric charge
$e$, (b) the magnetic field of a moving electron. (c) The magnetic field of
a static magnetic monopole with magnetic charge $g$, (d) the electric field
of a moving monopole. }
\label{fig1}
\end{figure}

This implies that electric and magnetic fields are symmetrical and
equivalent in this special case. Obviously, the electric-magnetic duality
symmetry is no longer true when $\rho _{e}\neq 0$ (or $\mathbf{J}_{e}\neq 0$%
). However, Dirac believed that the electromagnetic laws should have the
\textquotedblleft dual nature\textquotedblright\ under any circumstances, in
other words, the electric and magnetic fields may have a general intrinsic
symmetry and the Maxwell's equations of Eq. (\ref{maxwell}) are incomplete.
In 1931 \cite{dirac}, Dirac claimed that the mathematical introduction of
magnetic monopole (a basic unit of magnetic charge) into the Maxwell's
equations would lead to a symmetrical form of the Maxwell-Dirac equations
\begin{eqnarray}
\nabla \cdot \mathbf{E} &=&4\pi \rho _{e},  \notag \\
\nabla \cdot \mathbf{B} &=&4\pi \rho _{m},  \notag \\
\nabla \times \mathbf{E} &=&-\frac{1}{c}\frac{\partial \mathbf{B}}{\partial t%
}-\frac{4\pi }{c}\mathbf{J}_{m},  \notag \\
\nabla \times \mathbf{B} &=&\frac{1}{c}\frac{\partial \mathbf{E}}{\partial t}%
+\frac{4\pi }{c}\mathbf{J}_{e}.  \label{max_dirac}
\end{eqnarray}%
where $\rho _{m}$ is the magnetic charge density and $\mathbf{J}_{m}$ is the
magnetic current density. The above four equations would also be invariant
under the following transformations%
\begin{align}
\mathbf{E}& \rightarrow \mathbf{B};\mathbf{\quad B}\rightarrow -\mathbf{E},
\notag \\
\rho _{e}& \rightarrow \rho _{m};\quad \rho _{m}\rightarrow -\rho _{e},
\notag \\
\mathbf{J}_{e}& \rightarrow \mathbf{J}_{m};\quad \mathbf{J}_{m}\rightarrow -%
\mathbf{J}_{e}.  \label{duality1}
\end{align}

Dirac's monopole theory \cite{dirac} results into the following relation
between an electric charge $(e)$ and magnetic charge $(g)$

\begin{equation}
eg=\frac{hc}{4\pi }n=\frac{\hbar c}{2}n,\quad (n=1,2,3,\cdots )\text{ }
\label{quanta}
\end{equation}%
where $h$ is the Plank's constant, $\hbar =h/2\pi $ and $c$ is the speed of
light.

It should be pointed out that the magnetic monopole is merely a hypothetical
particle whose existence is postulated based on the duality symmetry. The
equivalence of the electric charge (electron) and the magnetic charge
(monopole) is explicitly shown in Fig. \ref{fig1}. Interestingly, Dirac
linked the magnetic monopoles with the quantization of electric charge by
Eq. (\ref{quanta}). Such appealing proposal exhilarated a number of
theoretical and experimental investigations since then. The numerous
attempts of experimental search for these magnetic monopoles at accelerators
and in cosmic rays have been done. And various techniques of detection in
the experiments to search for magnetic monopole have been developed, for
instance, the magnetometer SQUID. Unfortunately, up to now, no positive
evidence for its existence has been found. In theoretical physics, 't Hooft
\cite{hooft} pointed out that a unified gauge theory in which
electromagnetism is embedded in a semisimple gauge group would predict the
existence of the magnetic monopole as a soliton with spontaneous symmetry
breaking. Wu and Yang \cite{wu} first described magnetic monopoles in terms
of a principal of fiber bundle. Seiberg and Witten \cite{seiberg} developed
the famous magnetic monopole equations. The standard SU(5) model predicts
that the magnetic monopoles are extremely heavy with a mass at least 10$%
^{16} $ GeV/$c^{2}$ (the mass about $10^{15}$ protons), moreover, whose mass
is estimated to be even higher (up to the Planck mass of 10$^{19}$ GeV/$%
c^{2} $) by the Kaluza-Klein model.

What we are most concerned about is why no magnetic monopoles have been
detected after it had been hypothesized for 77 years. The experimental
status of monopoles had led Dirac to doubt his theory: \textquotedblleft I
am inclined now to believe that monopoles do not exist\textquotedblright\
\cite{dirac1}. In fact, several errors of the Dirac monopole theory have
been pointed out a long time ago \cite{zwanziger,weinberg,hagen}. In this
paper, we provide a solid argument that the hypothetical magnetic monopoles
aren't naturally real or the concept of magnetic monopole is only a
well-known particle (electron) of different representation.

\section{Magnet: electric charges or magnetic charges?}

\begin{figure}[tbp]
\begin{center}
\resizebox{1\columnwidth}{!}{
\includegraphics{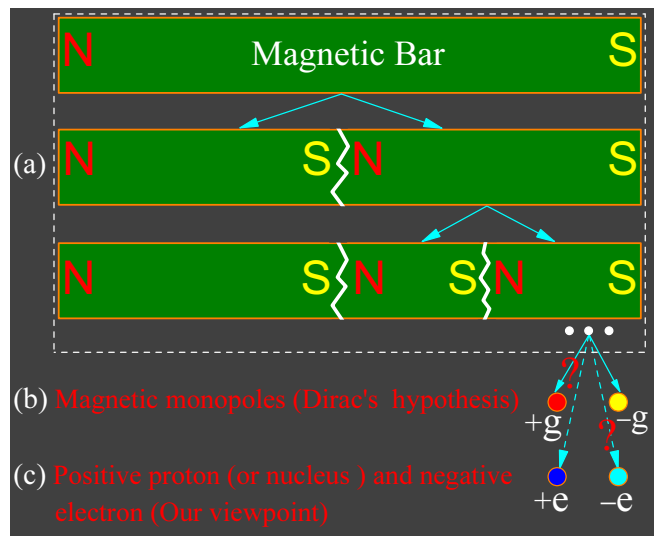}}
\end{center}
\caption{What is the most essential (smallest) component of a magnet? (a) As
a basic knowledge in electromagnetism, no matter how many times a bar magnet
is cut in half, there is always a north and a south pole. (b) Dirac put
forward the idea of the magnetic monopoles: the isolated $N$-pole ($+g$) and
the isolated $S$-pole ($-g$). (c) Our viewpoint is that the smallest magnet
is composed of a single proton (or nucleus) and a single electron. }
\label{fig2}
\end{figure}

It is well accepted that if a bar magnet is cut in half repeatedly, then
each half of the magnet becomes a separate magnet with its own north and
south poles, as shown in Fig. \ref{fig2} (a). Since the discovery of the
peculiar feature of the magnetic materials, many people are curious about
what it would look like if there was the smallest magnet, moreover, can the
smallest magnet be further isolated? According to Dirac's opinion, similar
to electric charges, there would have net magnetic charges (a magnet with
only one pole) in the universe, as shown in Fig. \ref{fig2} (b). Although
the assumption of the existence of the magnetic monopoles sounds
interesting, there are two fatal problems with this idea.

First, if the hypothetical magnetic particle is true in the natural world,
apparently, there should be plenty of the Dirac's magnetic monopoles inside
the permanent magnet materials. Hence, there is much more possibility to
detect the magnetic monopoles in the magnet materials than in the
accelerators and cosmic rays. In our opinion, no evidence for the monopole's
existence in the sources (magnet materials) for the magnetic monopoles may
indicate that monopoles do not exist at all.

Second, assuming there occurs a magnetic to non-magnetic transition in a
material, how and where are the magnetic monopoles going? If there are some
magnetic monopoles escaping from the material during the transition, as a
result, the mass of the material should be greatly reduced due to the
theoretical predication\ of the massive magnetic monopoles. Of course, if
one considers that all the magnetic monopoles still remain in the material
after the transition, then he has to explain what are the differences
between the monopole's states before and after the transition and why these
differences can not be experimentally detected.

From the viewpoint of the objectiveness and rationality of physics, when a
permanent magnet material is cut in the way of Fig. \ref{fig2}, there is no
doubt that ultimately we will obtain one positively charged proton and one
negatively charged electron, rather than the hypothetic magnetic monopoles,
as shown in Fig. \ref{fig2} (c). Now the question turn out to be
\textquotedblleft Can the real particles of proton and electron be used to
interpret the extremely common natural phenomenon described in Fig. \ref%
{fig2} (a)?\textquotedblright\ In the following sections, we will try to
answer this important question in a very intuitive way.

\section{Magnetic field or electric field?}

\begin{figure}[bp]
\begin{center}
\resizebox{1\columnwidth}{!}{
\includegraphics{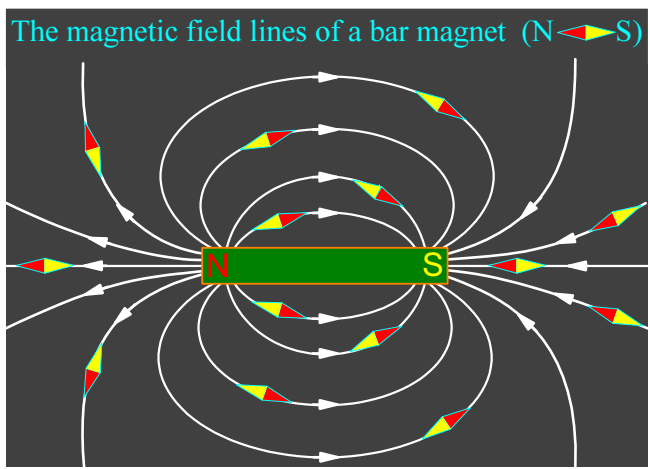}}
\end{center}
\caption{The static magnetic field of a bar magnet. The corresponding
magnetic field lines (the white curves) can be traced out with the use of
the compasses.}
\label{fig3}
\end{figure}

According to the traditional physics, the magnetic field lines of a bar
magnet form closed lines. The field direction is taken to be outward from
the North pole ($N$) and in to the South pole ($S$) of the magnet, as shown
in Fig. \ref{fig3}. The magnetic field lines, which can be traced out with
the use of the compasses (see also Fig. \ref{fig3}), are clearly more
concentrated around the two poles of the magnet. Basically, the space with a
denser magnetic field lines indicates a stronger magnetic field in that
region.

\begin{figure}[tp]
\begin{center}
\resizebox{1\columnwidth}{!}{
\includegraphics{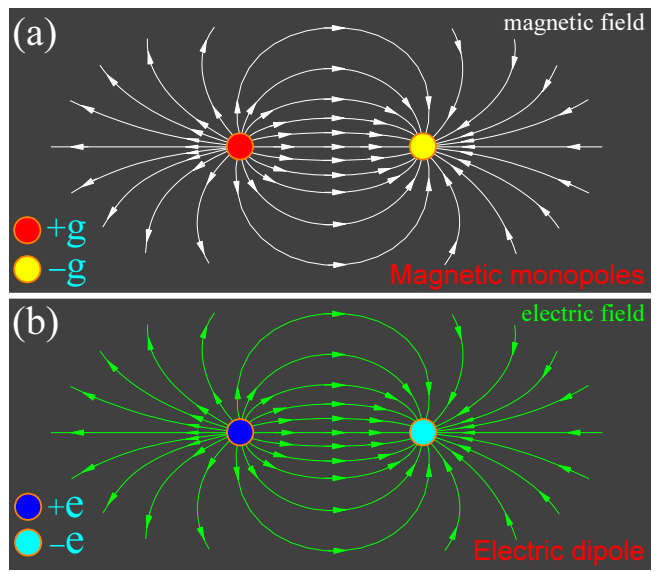}}
\end{center}
\caption{A comparison of the static magnetic field and the static electric
field. (a) The theoretical magnetic field lines (the white curves) of a pair
of hypothetical monopoles ($+g$ and $-g$), (b) the static electric field
lines (the green curves) of the simplest electric dipole consists of one
proton ($+e$) and one electron ($-e$). }
\label{fig4}
\end{figure}

If there really exist the monopoles with Dirac magnetic charges $+g$ and $-g$%
, then the magnetic field lines associated with a magnetic dipole can be
readily obtained, as shown in Fig. \ref{fig4}(a). As a comparison between
the magnetic field of the artificial magnetic dipole and the electric field
of the real electric dipole, in Fig. \ref{fig4}(b) we plot the well-known
electric field lines for the electric dipole. Similar to the case of the
magnetic dipole of Fig. \ref{fig4}(a), the electric field lines produced by
positive charge $+e$ will end in the negative charge $-e$. It is not
difficult to find that the two figures are identical. In fact, there is no
effective experimental means which can be used to distinguish between the
magnetic field of the so-called magnetic dipole and the electric field of
the electric dipole. In our opinion, the physical definition of the static
magnetic field is essentially an electric-dipole field. Namely, the widely
accepted physical concept of the static magnetic field most likely do not
exist in practice, it is therefore unnecessary to discuss the possible
existence of the magnetic monopoles (the sources of the static magnetic
field) in the nature.

\section{The nature of the static magnetic field}

\begin{figure}[tp]
\begin{center}
\resizebox{1\columnwidth}{!}{
\includegraphics{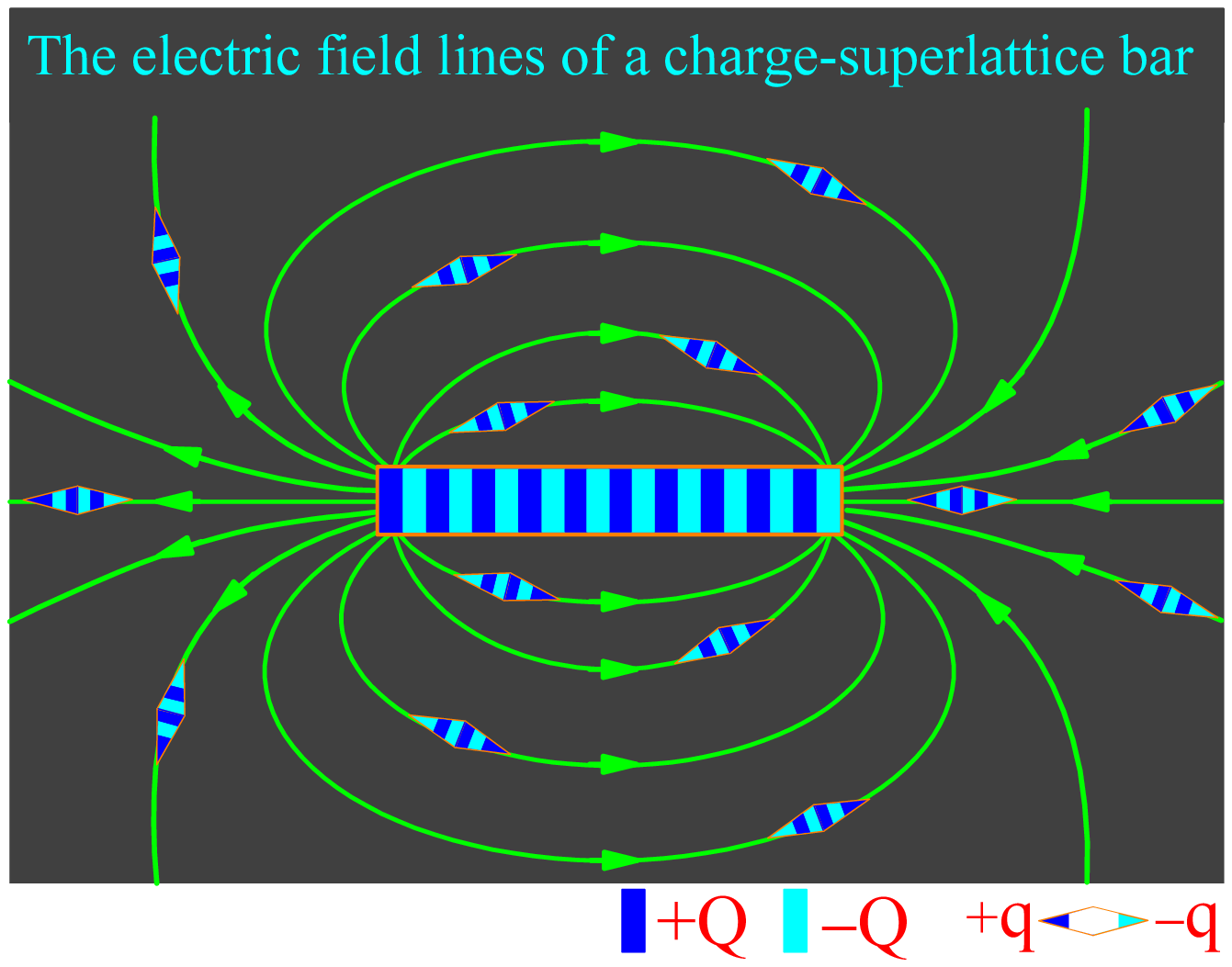}}
\end{center}
\caption{The static magnetic field of a bar magnet of Fig. \protect\ref{fig3}
can be perfectly generated by a electric bar of the periodically modulated
quasi-one-dimensional charge superlattice. }
\label{fig5}
\end{figure}

In order to make our argument of the nature of the static magnetic field
sounded, we try to design a bar \textquotedblleft magnet\textquotedblright\
and some compass needles with the positive and negative charges. As shown in
Fig. \ref{fig5}, the \textquotedblleft magnetic\textquotedblright\ bar and
the compasses have a superlattice structure comprising some pairs of layers
of positive and negative electric charges. With an appropriate
\textquotedblleft magnetic\textquotedblright\ bar (structure, size and
shape), the exactly same magnetic field lines of Fig. \ref{fig3} can be
generated by this periodic electric charge bar. At the same time these
electric compasses (see Fig. \ref{fig5}) can play the same role as the
magnetic compasses in Fig. \ref{fig3}. Now, the key question has been
whether such a periodically modulated charge structure can exist in real
magnetic systems.

\begin{figure}[tp]
\begin{center}
\resizebox{1\columnwidth}{!}{
\includegraphics{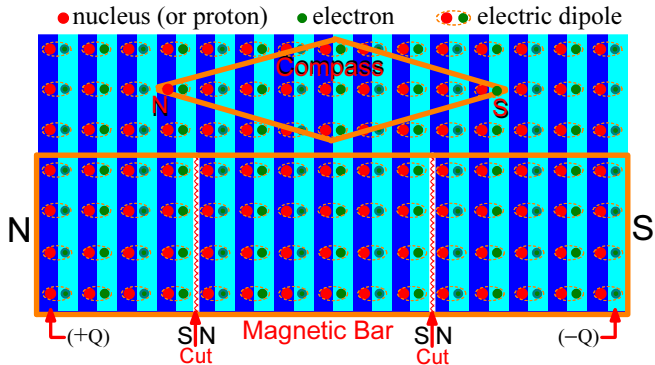}}
\end{center}
\caption{The superlattice of Fig. \protect\ref{fig5} with the alternate
positive and negative charges periodic structure can be expected in some
transition metals where the nucleus and the corresponding localized $d$%
-electron form a electric dipole. All the magnetic properties can be well
explained by this picture, as shown in the figure.}
\label{fig6}
\end{figure}

To the best of our knowledge, why some elements have the so-called intrinsic
magnetic property (IMP) is still an unsolved problem in condensed matter
physics. In accordance with the picture of Fig. \ref{fig5}, it now seems
more clear that, to exhibit the IMP, a quasi-one-dimensional periodic
structure of the positive and negative charges must be naturally formed in
the elements (or materials). In some transition metals with the IMP, it is
reasonable to assume that each atom contains one nucleus carrying one net
positive basic charge and one localized $d$-electron carrying one negative
basic charge that form a smallest electric dipole. As shown in Fig. \ref%
{fig6}, the nuclear and electrons can organize into a electric-dipole
crystal with the alternate positive and negative charges periodic structure.
With the help of this figure, all the so-called magnetic properties
occurring in nature can be well explained. For example, when a bar of the
electric-dipole is cut arbitrarily across the axis direction, each piece
always has its own positive charge end (or $N$-pole) and negative charge end
(or the $S$-pole), as indicated in Fig. \ref{fig6}.

\section{Conclusion}

In this paper, on one hand, we have studied the possibility of the existence
of the Dirac's magnetic monopoles, on the other hand, we have attempted to
uncover the physical nature of the static magnetic field generated by a bar
magnet. It was shown clearly that the concept of the massive magnetic
monopoles is physically untrue. The hypothetical particle is likely to be
the well-known electron. This result indicates that any attempts to search
for the magnetic monopole in the universe will be proved to be in vain. We
have found that the traditional static magnetic field of a bar magnet, in
fact, is the static electric field of the periodically quasi-one-dimensional
electric-dipole superlattice. It seems that we had misdefined the static
electric field of the electric-dipole lattice as the magnetic field of the
magnet (or the magnetic monopoles). Interestingly, this new concept of
periodic structure of the positive and negative charges may proved to be
true in some transition metals with the intrinsic magnetic property. Further
related research is being conducted.

\end{document}